\documentstyle[nato]{crckapb}

\def\a{\alpha}
\def\b{\beta}
\def\c{\chi}
\def\d{\delta}
\def\e{\epsilon}                
\def\f{\phi}                    
\def\g{\gamma}

\def\j{\psi}

\def\l{\lambda}
\def\m{\mu}
\def\n{\nu}

\def\p{\pi}                     

\def\J{\Psi}
\def\L{\Lambda}
\def\O{\Omega}

\def\S{\Sigma}

\def\cp{{\cal P}}

\def\car{{\cal R}}

\def\bo{\raisebox{-.4ex}{\large$\Box$}}                 
\def\cbo{{\,\raise-.15ex\Sc [\,}}                       

\def\sl#1{\rlap{\hbox{$\mskip 1 mu /$}}#1}      
\def\bra#1{\Big\langle #1\Big|}                 
\def\ket#1{\Big| #1\Big\rangle}                 

\def\svev#1{\left\langle #1\right\rangle}       

\def\ddt#1{{\buildrel {\hbox{\LARGE .\kern-2pt.}} \over {#1}}}

\def\beq{\begin{equation}}
\def\eeq{\end{equation}}
\def\bqry{\begin{eqnarray}}
\def\eqry{\end{eqnarray}}

\def\beqn#1{ \renewcommand{\theequation}{#1}
             \begin{eqnarray} }
\def\eeqn{ \renewcommand{\theequation}{\arabic{equation}}
           \end{eqnarray} }

\def\beqr#1{ \setcounter{equation}{#1}
             \begin{eqnarray} }

\def\eeqr{\end{eqnarray}}
\def\NON{\nonumber\\}
\def\beqrabc#1{ \setcounter{equation}{0}
                \renewcommand{\theequation}{#1\alph{equation}}
                \begin{eqnarray} }
\def\beqrn#1#2{ \setcounter{equation}{#2}
                \renewcommand{\theequation}{#1.\arabic{equation}}
                \begin{eqnarray} }
\def\seeq#1{eq.~(\ref{#1})}

\def\seeqs#1{eqs.~(\ref{#1})}

\def\seneq#1{~(\ref{#1})}

\def\NPB#1{Nucl. Phys. {\bf B#1}}
\def\NPBP#1{Nucl. Phys. (Proc. Suppl.) {\bf B#1}}
\def\PLB#1{Phys. Lett. {\bf B#1}}
\def\PRD#1{Phys. Rev. {\bf D#1}}

\def\PRL#1{Phys. Rev. Lett. {\bf #1}}

\def\sstyle{\scriptstyle}

\def\rhs{\mbox{r.h.s.} }

\def\ie{\mbox{\it i.e.} }
\def\eg{\mbox{e.g.} }

\def\geqx{\,\raisebox{-1.0ex}{$\stackrel{\textstyle >}{\sim}$}\,}

\def\frac#1#2{ {\sstyle {#1\over #2} } }
\def\det#1{{\rm det}\left(#1\right)}

\def\tr{{\rm tr}\,}

\def\Re{{\rm Re\,}}

\def\det{{\rm det\,}}

\def\GeV{\;{\rm GeV}}
\def\csv{\mbox{$\c$SV} }
\def\csvp{\mbox{$\c$SV}}
\def\hatT{\hat{T}}
\def\tT{\tilde{T}}
\def\qeff{q_{\rm eff}}
\def\qpt{q_{\rm pt}}
\def\npf{N_{\rm pf}}
\def\rcite#1{ref.~\cite{#1}}

\begin{opening}
\title{BETTER DOMAIN-WALL FERMIONS}

\author{Yigal Shamir}
\institute{School of Physics and Astronomy\\
Beverly and Raymond Sackler Faculty of Exact Sciences\\
Tel-Aviv University, Ramat~Aviv 69978, Israel}

\end{opening}

\begin{document}

\begin{abstract}
We discuss two modifications of domain-wall fermions,
aimed to reduce the chiral-symmetry violations presently
encountered in numerical simulations\footnote{
contribution to workshop 
``Lattice Fermions and Structure of the Vacuum" (October 1999, Dubna, Russia)
}.
\end{abstract}

\section{INTRODUCTION}

Domain-wall fermions (DWF) live on a five-dimensional lattice~\cite{k,nn,fs}.
The fifth coordinate, usually denoted $s$, is introduced because
chiral-symmetry violations (\csvp) decrease with increasing size
of the fifth dimension, $N_s$.

DWF are rapidly becoming
a standard method for lattice QCD simulations (for a review see \rcite{rev}).
While showing agreement with other methods for quantities
such as the strange-quark mass and $B_K$~\cite{msbk},
a recently reported first result for $\e'/\e$ \cite{eprme} signals that DWF
may be employed to compute the Standard-Model predictions
in cases where the traditional (Wilson, staggered) fermion methods 
have so far been unsuccessful.
DWF have also been employed for the study of
thermodynamics, topology and the chiral condensate~\cite{thermo}.
We note that the four-dimensional overlap-Dirac operator~\cite{nn,nop}
arises in a special limit of DWF~\cite{kiku}, 
hence results pertaining to this operator have relevance for DWF too.

When the inverse lattice spacing is
$a^{-1} \sim 2 \GeV$ (quenched $\b \sim 6.0$) \csv are found to be
at the few percent level for $N_s \sim 15-20$ 
for most quantities~\cite{qsim,msbk,eprme,qual}. 
However, at stronger coupling,
$a^{-1} \sim 1 \GeV$, the violations are much bigger~\cite{thermo,qual},
and sometimes even $N_s \sim 50$ does not seem to be enough.
Since the cost of simulations grows linearly with $N_s$,
an obvious question is whether fiddling with the DWF action
may allow us to decrease $N_s$ while maintaining (or even improving)
the quality of chiral symmetry.

In this report we describe two modifications of DWF,
aimed to reduce the presently encountered \csvp.
The first method, discussed in Sect.~2, is applicable
in dynamical simulations~\cite{redch}.
This method is based on non-perturbative considerations.
The second method (Sect.~3) is based on perturbative considerations,
and is applicable in both quenched and dynamical simulations~\cite{w2}.

\section{A MODIFIED PSEUDO-FERMION ACTION}

The DWF determinant can be written as
\beq
  \det(D_F) = (\m_{F})^{N_s} \times (\mbox{finite factor}) \:.
\label{bulk}
\eeq
Here $D_F$ is the five-dimensional DWF matrix
(we use the notation of ref.~\cite{redch}
with lattice spacings $\,a=a_5=1$).
The ``finite factor" has a convergent $N_s\to\infty$ limit.
It contains the physics of the quark field,
whose two chiral components live on opposite four-dimensional boundaries
of the lattice.
The bulk factor $(\m_{F})^{N_s}$ arises because of the presence of extra
$N_s - 1$ four-dimensional fermions with masses of the order of the cutoff.
This undesirable factor
is canceled by introducing opposite-statistics Pseudo Fermion (PF)
fields~\cite{nn,frsl}. In the so-called second order formulation,
the PF action is
\beq
  S_{\rm pf} = \sum_{xs,ys'}
  \f^\dagger_{xs}\, (D_{\rm pf} D_{\rm pf}^\dagger)_{xs,ys'}\; \f_{ys'} \:,
\label{Spf}
\eeq
where the PF lattice has only $\npf=N_s/2$ sites in the fifth direction.
The standard choice is $D_{\rm pf}=D_F(m=1)$. 
The parameter $m$ controls the coupling between the two boundaries
of the five-dimensional lattice
(hence also between the left and right components of the quark field),
and is proportional to the bare quark mass for $m\ll 1$.
The choice $m=1$ corresponds to
anti-periodic boundary conditions, and implies the absence
of any light PF states.

Aiming to reduce \csv we consider a modified PF action
\beq
  S_{\rm pf}(c) = \sum_{xs,ys'}
  \f^\dagger_{xs}
  \left[ (D_{\rm pf} D_{\rm pf}^\dagger)_{xs,ys'}
         + c\, \cp_x \, \d_{x,y}\,\d_{s,s'}
  \right]  \f_{ys'}\:,
\label{Smod}
\eeq
where
$\cp_x =
\sum_{\m < \n} \Re \tr (I - U_{x\m}U_{x+\hat\m,\n}
          U^\dagger_{x+\hat\n,\m}U^\dagger_{x\n})
$
is the plaquette-action density at the point $x$,
and $c$ is a coupling constant.

In Subsect.~2.1 below we discuss what we believe to be
the primary source of \csv at strong coupling.
In Subsect.~2.2 we explain how the modified PF action should
help reducing \csvp. Our considerations also suggest
how to look for the parameters' range where the new PF action
may be effective. The new term in \seeq{Smod}
does not change the sparseness of the (2nd order) PF matrix.
Therefore the extra cost of simulating it is hopefully not large.
(One should give thought to finding a good pre-conditioning.)

\subsection{the problem}

It was first observed by Narayanan and Neuberger~\cite{nn}
that the DWF determinant can be compactly expressed
in terms of a (second-quantized) transfer matrix $\hatT$
that describes single-site hopping in the unphysical $s$-direction.
The $\hatT$-formulation is especially powerful for
DWF, because the same gauge field couples to the fermions
on every $s$-slice, and so $\hatT$ is independent of $s$.

As expected from their good chiral properties,
DWF have (on a finite lattice, almost) exact zero modes
in an instanton background.
As a topologically non-trivial gauge field is gradually
turned on, at some point level-crossing occurs
in the spectrum of the background-field dependent 
hamiltonian $\hat{H}=\log\hatT$,
and an initially-excited state becomes 
the new ground state~\cite{nn,scri1,smth}.
Beyond the crossing point the axial U(1) charge is not conserved,
as required by the anomaly. At the crossing point,
one of the eigenvalues of $T$, the corresponding
first-quantized transfer matrix, is equal to one.
Two comments are in order.
1) An eigenvector of $T$ with eigenvalue one exists if and only if
there exists a zero eigenvector of the hermitian Wilson-Dirac operator
(this extends to a mapping between all eigenvectors of $T$
and of the hermitian Wilson-Dirac operator~\cite{redch}).
2) Since the mass of the Wilson-Dirac operator is supercritical,
and is far from the critical mass $M_c$,
the above zero eigenvectors have nothing to do with continuum physics.
They are highly localized states whose support typically extends
over only one or two lattice spacings~\cite{scri2}.
Moreover, one can rigorously prove the absence of such zero eigenvectors
if the plaquette-action density is everywhere smaller than a
certain constant~\cite{bound}.

A near-unity eigenvalue (NUEV) of $T$ implies
little or no suppression of $s$-correlations
for quark states that have a finite overlap with
the corresponding (four-dimensional) eigenvector.
Strong correlations across the $s$-direction also mean
strong lattice-artifact \csv for (non-singlet) axial currents~\cite{fs}.
As follows from the picture of topology
changing developed in ref.~\cite{nn}, NUEVs
cannot be completely avoided. This said,
the {\it spectral density} of NUEVs is a very
sensitive function of the lattice parameters.
The dependence of the spectral density on $\b$ was investigated
in ref.~\cite{scri3} for quenched ensembles.
The results (see in particular fig.~5 therein)
confirm that NUEVs are extremely rare for $\b \ge 6.0$. 
Practically, this implies an exponential fall-off
of all correlations in the $s$-direction.
However, at stronger coupling there is a dramatic rise in the
density of NUEVs. In this case \csv  decay very slowly  
and are significant even for very large $N_s$ \cite{qual}. 
In fact, it was recently proved within the strong-coupling expansion 
that explicit \csv persists in the limit $N_s \to \infty$,
and that the symmetry group of the effective hamiltonian 
is identical to that of naive fermions with a mass term~\cite{rbbs}.

What is the role of NUEVs in a dynamical simulation?
As we explain in Subsect.~2.2 below, 
to every NUEV of $T$ corresponds a {\it family} 
of small eigenvalues of the DWF operator.
If ``small'' means, say, $O(\L_{QCD})$,
the number of small eigenvalues of $D_F$ is $O(N_s\,\L_{QCD})$.
Why, then, are the relevant ``bad'' gauge-field configurations
not suppressed for $N_s \gg 1$? 
The same set of small eigenvalues occurs also in the PF determinant
which sits in the denominator, and the two sets
simply cancel each other.
This is a characteristic feature of small eigenvalues 
of $D_F$ which are related to NUEVs of $T$. In contrast, 
the number of DWF (near) zero modes found in an instanton background
is as required by the anomaly (not proportional to $N_s$)
and with no matching PF zero modes.

One can say that ``bad''  gauge-field configurations
are not suppressed because the PF determinant follows
the DWF one too closely! However, there is no need
to maintain this tight relation far from the continuum limit.
The purpose of the new PF action \seeq{Smod}
is to break this relation in a controlled way,
such that configurations with NUEVs will be suppressed.

\subsection{a possible solution}

We now turn to the analysis of the modified PF action.
(The PF action\seneq{Smod} was first introduced in ref.~\cite{redch}.
The below discussion is largely new. While we try to make this report
self-contained, some familiarity with ref.~\cite{redch}
should help in keeping track of what follows.) The crucial step
is to rewrite the PF determinant as a product,
${\rm det}^{\npf}(B)\; {\rm det}(\O(c))$, where
\beq
 \O(c)_{xs,ys'} = - \bo_{s,s'}\, \d_{x,y} + \d_{s,s'}\, \O_4(c)_{x,y}
                  + \mbox{boundary terms}\,,
\label{omega}
\eeq
\beq
 \O_4(c) = \g_5 B^{-1/2}
  \left[ D_{\rm pf} D_{\rm pf}^\dagger + c\, \cp \right] \g_5 B^{-1/2} \,.
\label{o4}
\eeq
Here $D$ is the massless Wilson-Dirac operator,
$\cp$ is the diagonal matrix with entries $\d_{x,y}\cp_x$,
and $B=1-M+W$ where $W$ is the Wilson term and $M$
is the five-dimensional mass, or ``domain-wall height''.
(Both $M$ and $W$ are assumed to be
positive; because of the relative minus sign
this corresponds to supercritical Wilson fermions.)
The free nearest-neighbor laplacian for the $s$-coordinate
is $\bo_{s,s'} = \d_{s,s'+1} + \d_{s,s'-1} - 2 \d_{s,s'}$.
The boundary terms in \seeq{omega} 
are zero unless $s$ or $s'$ are equal to 1 or $\npf$.
We will be interested in computing how the Boltzmann weight
varies as a function of $c$, \ie in the ratio
\beq
  \exp(-\car(c)) \equiv {\rm det}(\O(0))/{\rm det}(\O(c)) \,.
\label{r}
\eeq
Since the (PF) action is local,
the free energy is dominated by a bulk term
that grows linearly with $\npf$.
We will only be interested in the way the PF bulk factor changes.
Writing $\car(c) = \npf\, \car_1(c) + \car_0(c) + O(1/\npf)$
we observe that that $\car_1(c)$ is independent of
the boundary terms in \seeq{omega} (this can be proved using
the determinant formulae of \rcite{redch}).
Denote by $\O'(c)$ the operator that coincides with $\O(c)$,
except that the boundary terms are absent (i.e. $\O'(c)$
lives on a periodic lattice in the $s$-direction).
The eigenvectors of $\O'(c)$ are given by $e^{i s p_5}\, \J_\l(x)$ 
where $e^{i s p_5}$ is a one-dimensional
plane wave and $\J_\l(x)$ is an eigenvector of $\O_4(c)$
with eigenvalue $\l(c)$ ($\O_4(c)$ is positive, hence $\l(c) \ge 0$ always).

The eigenvalues of $\O'(c)$ are given by $\l(c) + 2(1-\cos(p_5))$.
Consider first the situation at $c=0$.
If the transfer matrix has a unity eigenvalue,
$\O_4(0)$ has a zero eigenvector $\j_0$,
and therefore $\O'(0)$ has a gap-less family of eigenvalues
given by $2(1-\cos(p_5))$. This explains the statement made
in the previous subsection about the number of small DWF and PF
eigenvalues for ``bad'' configurations.
Projected \cite{redch} onto the four-dimensional eigenvector $\j_0$,
the DWF propagator 
$G(s,s') = \sum_{p_5} e^{ip_5(s-s')}\, (1-\cos(p_5))^{-1}$
does {\it not} fall exponentially with the separation $|s-s'|$, 
leading to unsuppressed correlations across the $s$-direction.

In computing $\car_1(c)$ we will simplify things by replacing
the lattice cutoff for the $s$-direction with a sharp momentum
cutoff. Since the modification of the PF action is intended to be small,
it should affect mainly the small eigenvalues,
and changing the cutoff scheme should be harmless.
As a second simplification we will use first-order perturbation
theory to determine how the eigenvalues of $\O_4(c)$ vary.
One has $\l(c) \approx \l_0 + c\, \l_1$ where
\beq
  \l_1 = \bra{\J_{\l_0}} B^{-1/2}\, \cp\, B^{-1/2} \ket{\J_{\l_0}} \,,
\label{l1}
\eeq
(note that $\l_1 \ge 0$).
Replacing the fifth-coordinate momentum sum by
an integral (which is allowed for large $\npf$)
yields the following result
\bqry
  \car_1(c) & = & \sum_{\l_0} \int {dp_5\over 2\p}\,
  \left( \log(p_5^2 + \l_0 + c\, \l_1) - \log(p_5^2 + \l_0) \right)
\NON
  & = & \sum_{\l_0}
  \left( \sqrt{\l_0 + c\, \l_1} - \sqrt{\l_0}\, \right) \,.
\label{r1}
\eqry
We have sent the momentum cutoff to infinity.
The result is finite, and this justifies
our treatment of the $s$-coordinate's ultra-violet cutoff.

We now wish to distinguish two, qualitatively different, cases
depending on the gauge-field background. First assume that there is a gap in
the spectrum of $\O_4(0)$, namely $\l_0\ge \l_{\rm min} > 0$
for all $\l_0$ where by assumption $\l_{\rm min}=O(1)$. 
Taking $c \ll \l_{\rm min}$ (note that $\l_1=O(1)$ by \seeq{l1}) 
and expanding the first term in \seeq{r1} we have
\beq
  \car_1(c) = c \sum_{\l_0} { \l_1 \over 2\sqrt{\l_0}} \,.
\label{gap}
\eeq
What is the physical significance of this result?
We can compute $\car_1(c)$ also by varying the PF partition
function (defined with periodic boundary conditions)
with respect to $c$. This yields
\beq
  \car_1(c) = c \sum_x \cp_x \svev{\f^\dagger_{xs}\f_{xs}} \,,
\label{HF}
\eeq
(for any fixed $s$).
One can show that \seeqs{gap} and\seneq{HF} agree for $c \ll \l_{\rm min}$.
Invoking translation invariance \ie replacing 
$\langle \f^\dagger_{xs}\f_{xs} \rangle$
by its four-dimensional average
$V_4^{-1} \sum_x \langle \f^\dagger_{xs}\f_{xs} \rangle$
tells us that the effect of the modified PF action
is an additive renormalization of the inverse bare coupling
\beq
  \b \to \b + c\, \npf V_4^{-1} \sum_x \svev{\f^\dagger_{xs}\f_{xs}} \,.
\label{dbeta}
\eeq
An estimate of the expectation value in \seeq{dbeta}
may be obtained using standard weak-coupling perturbation theory.

The second possibility is that the  spectrum of $\O_4(0)$ is gap-less.
For the low eigenvalues of $\O_4(0)$ we can no longer use \seeq{gap}.
Instead, for any $\l_0\ll c$ one has
\beq
  \sqrt{\l_0 + c\, \l_1} - \sqrt{\l_0} \approx \sqrt{c\, \l_1} \,.
\label{nogap}
\eeq
We see that, in the absence of a gap,
$\car_1(c)$ is not a linear function of $c$.
This happens because of the singularity that ${\rm det}(\O(0))$
develops when an eigenvalue $\l_0$ tends to zero. For small $c$,
the \rhs of \seeq{nogap} is much bigger than $c\,\l_1/\sqrt{\l_{\rm min}}$.
Thus, as desired, no-gap configurations are relatively suppressed!

In practice one can try to implement this new PF action
along the following lines. Suppose that we wish to reduce \csv
in a dynamical simulation with given $\b = \b_0$.
We start instead with a smaller value, 
say in the range $\b \sim 0.8 \b_0$ -- $0.9 \b_0$,
and increase $c$ until the measured lattice spacing
agrees with the one previously obtained for $\b=\b_0$.
At this point one should have a coupling constant $c$ of order $O(1/N_s)$.
(When varying $N_s$ it is advisable to keep the product $c\, N_s$ invariant
by taking $c=c_1/N_s$ with fixed $c_1$.)
For configurations with a gap the Boltzmann weight should 
now be roughly the same as before, because the reduction in the explicit $\b$
is matched by the effective increase in $\b$ coming
from the PF action. On the other hand, no-gap configurations
should now have a much smaller Boltzmann weight,
since the suppression provided by the new PF action wins over
the effect of working at smaller $\b$.
Thus, if there is a ``window" where all the simplifications made
above approximately hold,
this method should lead to significantly reduced \csvp.

\subsection{addendum}

We close this section with two comments. 
The favorable value of $M$ is known to be $M \sim 1.8$.
This implies that $B=1-M+W$ is not necessarily a positive operator
and, consequently, that the transfer matrix is not necessarily 
positive either. 
In practice, it turns out that the positivity of $B$ is preserved. 
For example, for an ensemble of quenched $\b=5.7$ configurations, 
the spectrum of $B$ was found to be bounded away from
zero while $M < 2$, so the Wilson term has
an effective lower bound of around one~\cite{comm}. 
A similar shift is found for $M_c$, the critical Wilson-fermion mass, 
which is also compatible with 
the mean-field and one-loop predictions~\cite{at}.
We thus have a nicely consistent picture of several quantities
all having the same origin -- the rise in the effective lower
bound on $W$ away from the continuum limit.

The second comment corrects a misleading statement made in \rcite{redch}.
There, concern was expressed that a slightly different transfer matrix
denoted $\tT$ could have complex eigenvalues.
This actually cannot happen~\cite{pv} because the transfer
matrices $\tT$ and $T$ have the same spectrum.
When $B$ is negative, $T$ may indeed have negative eigenvalues,
but cannot have complex eigenvalues since its definition 
involves only $B$ and its inverse, but not $\sqrt{B}$.
(Having negative eigenvalues for $T$ could also be annoying,
since when $\log(T)$ does not qualify as a hamiltonian
there could appear new unwanted effects;
this unpleasant situation, too, seem to have little relevance
because of the reasons explained in the previous paragraph.)

\section{THE FOUR-DIMENSIONAL PART OF THE ACTION}

In this section we propose a different modification of the DWF action,
which may be used in both quenched and dynamical simulations
(possibly in conjunction with the modified PF action in the latter
case; the full details will be reported separately~\cite{w2}).
We begin by noting that, for a semi-infinite $s$-coordinate,
the quark state with zero four-momentum attached to the $s=1$ boundary
falls like $\c(s) \propto |1-M|^s$.
(This should not be confused with the ``quark field'' introduced in \rcite{fs}.
The latter is an interpolating field for quark states which by
definition is restricted to the boundary layers.)
On a finite lattice, the mixing of the quark states from the two boundaries
results in a residual bare quark mass $m_{\rm res}=O((1-M)^{N_s})$,
even if the explicit bare mass $m$ is zero~\cite{fs,pvns}.

In a simulation, the measured pion-mass squared 
will be proportional to the sum $m+m_{\rm res}$ 
(as long as both $m$ and $m_{\rm res}$ are small) 
but $m_{\rm res}$ is much larger
than suggested by the tree-level result.
Assuming that low momentum quark states behave in simulations
like $\c(s) \sim \qeff^s$, one expects $m_{\rm res} \propto \qeff^{N_s}$. 
Thus, we can extract $\qeff$
from the $N_s$-dependence of the pion-mass squared
(or of the anomalous term in the relevant Ward identity).
Results at quenched $\b=6$ \cite{qsim} suggest a value $\qeff
\sim 0.8$. This value is very large, if we remember that
$\qeff = 1$ means {\it no} exponential fall-off.

In the free DWF case, one has $q_0=|1-M| \to 0$ for $M \to 1$.
Also in a mean-field approximation~\cite{at} one finds a similar result
$q_{\rm mf}=|1+\d M - M|$. If mean-field was a good description,
letting $M \to 1+\d M$ would give $\qeff \ll 1$ in simulations.
Since in practice $\qeff$ is much closer to one than it is to
zero, we conclude that the effective value of $\d M$ varies considerably
over different configurations and, in fact,
over different spacetime regions of the same configuration.

We see that tree-level or mean-field approximations
fail to describe a key feature of the quark's wave function.
One reason for this failure, namely NUEVs, was already discussed in the
previous section. However, we believe that NUEVs cannot be the whole
story. A large density of NUEVs implies no exponential fall-off at all,
and a non-zero residual mass even for $N_s \geqx 50$.
On the other hand, for quenched $\b \ge 6.0$ the exponential fall-off
seems to be there, albeit with a pretty large $\qeff$.
As we will now explain, we believe that the large value of $\qeff$
is essentially generated by {\it perturbative} fluctuations
of the gauge field
(this is also supported by the results of \rcite{pvns,smth}).
Below, we use the one-loop expression for the DWF self-energy derived
in~\rcite{at} (for one-loop results for composite operators
see~\rcite{pt}).

\subsection{the perturbative wave function}

Let us assume $m=0$ and $N_s \gg 1$, so that the tree-level value of
$m_{\rm res}$ is negligible.
For small four-momentum $p_\m$
the free DWF propagator in the vicinity of the $s=1$ boundary is
\beq
  G_0(p_\m;s,s') = \c_0(s)\,{1\over \sl{p}}\,\c_0(s')\, P_L + {\rm Reg.}
\label{Gfree}
\eeq
where $\c_0(s) = |1-M|^s$ is again the tree-level
quark's wave function, $P_{L} = {1\over 2}(1 - \g_5)$
and ``Reg'' stands for a regular function of $p_\m$.

At the one-loop level,
the leading quantum effect is the additive renormalization
$M \to M + \d M$ mentioned above~\cite{at}.
This effect comes from a tadpole self-energy diagram,
and must be treated non-perturbatively (``tadpole improvement'').
Having done so, we obtain the 
resummed one-loop propagator~\cite{norman}
\beq
  G^{(1)}(p_\m;s,s') = \c_1(s)\, {1\over Z\sl{p}}\, \c_1(s')\, P_L
  + {\rm Reg.}
\label{Gone}
\eeq
where $Z=1+O(g^2)$ is a standard wave function renormalization factor.
At the one-loop level we find the wave function
\beq
  \c_1(s) = |1+\d M - M|^s + g^2\, \d\c_1(s) \,,
\label{wf1}
\eeq
\beq
  \d \c_1(s) \propto {\rm tr}\, P_L\,\S(k_\m=0;s'=1,s) \, 
          \propto s^{-2}\, \left({1\over 2}\right)^s \,,
\label{half}
\eeq
where $\S(k;s',s)$ is the DWF self-energy
coming from the ``setting sun'' diagram~\cite{at}
(${\rm tr}\, P_L\,\S(k;s',s)$ has a smooth $k_\m \to 0$ limit).
The self-energy can be written as
\beq
   {\rm tr}\, P_L \, \S(k;1,s) =
  \int_{B.Z.} d^4p \; h(k,p) \, \exp(- s\, \a(p)) \,.
\label{selfe}
\eeq
The exponential containing the $s$-dependence
arises from the internal fermion line with four-momentum $p_\m$.
All other terms are represented by $h(p,k)$.

In the limit $s \gg 1$ one can employ a saddle-point approximation
because ${\rm tr}\, P_L \, \S(k;1,s)$ is dominated by the maximum 
of $\exp(-\a(p))$ over the Brillouin zone. 
The exponents $\a(p)$ are determined by the tree-level
DWF action~\cite{nn,fs}.
For ordinary DWF ${\rm max}\{\exp(-\a)\}=0.5$ at $M=1$. This maximum
is obtained at the four points $P_\p=\{(\p,0,0,0),(0,\p,0,0),\ldots\}$.
(The values for any $0<M<2$ were computed in \rcite{ktv}.) 
The $s^{-2}$ factor in \seeq{half} arises because the (gaussian)
integration is four-dimensional.

Within tadpole-improved perturbation theory and
assuming $M$ has been tuned to $1+\d M$,
the $s$-dependence of the quark's wave function 
is governed by $\d\c_1(s)$. Ignoring the pre-exponential factor
in \seeq{half}, we thus find to one-loop order the {\it universal} result 
$\qpt ={\rm max}\{\exp(-\a)\}= 0.5$.
Physically, what the result
means is that propagation in the $s$-direction is dominated by 
the four states belonging to $P_\p$.
In perturbation theory, the wave functions of
all quark states have roughly the same $s$-dependence, namely $\qpt^s$, 
because they all communicate with the states in $P_\p$.

\subsection{new four-dimensional terms}

If we want to improve the fall-off of the quark's wave function,
we must look for DWF actions where ${\rm max}\{\exp(-\a)\}$
is smaller. We stress that, since it is determined 
by the tree-level DWF action, $\a(p)$ is unchanged if 
one employs ``fat links" or an improved gauge action.
We will first consider here DWF actions where $\bar\j_x$ 
couples to both $\j_{x\pm\hat\m}$
and $\j_{x\pm 2\hat\m}$, \ie we allow for next-nearest neighbors,
but only in the same direction.
For the $s$-coordinate we retain the same nearest-neighbor
coupling as with ordinary DWF. 

The four-dimensional part
of the usual DWF action contains the familiar Wilson-Dirac operator.
Here we replace it by a new operator,
whose tree-level momentum-space form is
\beq
 D_{23} = \sum_\m \g_\m f_3(p_\m) + r\, W_2 - M \,,
\label{d23}
\eeq
\beq
  f_3(p_\m) =  \sin(p_\m)\left[1+c_3(1-\cos(p_\m))\right] \,,
\label{f3}
\eeq
\beq
  W_n = \sum_\m (1-\cos(p_\m))^n \,,
\label{wn}
\eeq
(the standard Wilson-Dirac case corresponds to $c_3=0$, $n=1$).
For $M=1$, following \rcite{nn,fs} the exponents are determined by
\beq
  2\, {\rm cosh}(\a) = {1 + r^2\, W_2^2 + \sum_\m f_3^2(p_\m)\over r\, W_2}
\label{cosh}
\eeq
where by convention  $\a>0$.
In ref.~\cite{w2} we analyze the mathematical structure 
of the minima of \seeq{cosh}
and explain why taking $n=2$ and $c_3>0$ raises
the global minimum of ${\rm cosh}(\a)$
which, in turn, lowers the global maximum of $\exp(-\a)$.
A few examples are given in Table~1. The values in the last column
should be compared with ${\rm max}\{\exp(-\a)\}=0.5$ of ordinary DWF.

\begin{table}[htb]
\begin{center}
\caption{${\rm max}\{\exp(-\a)\}$ for the operator $D_{23}$
and for various values of
$c_3$ and $r$ (see text for the definitions).
The second column gives the first two terms
in the expansion of $f_3(p)$. For each $c_3$ we looked for
the $r$-value where ${\rm max}\{\exp(-\a)\}$ is smallest.
}
\begin{tabular}{clccc}
\hline
$c_3$ &  $\;\;f_3(p)$ &   $r$ &   
$ 2 {\rm min}\{\cosh(\a)\}$    &    ${\rm max}\{\exp(-\a)\}$ \\
\hline
0            &   $p - {1\over 6} p^3$ &  1.46 &  2.83 &  0.414  \\
${1\over 3}$ &   $p                 $ &  1.14 &  3.40 &  0.326  \\
${2\over 3}$ &   $p + {1\over 6} p^3$ &  1.19 &  4.09 &  0.261  \\
${4\over 3}$ &   $p + {1\over 2} p^3$ &  1.45 &  5.62 &  0.184  \\
${7\over 3}$ &   $p + p^3$            &  1.98 &  8.06 &  0.126  \\
\hline
\end{tabular}
\end{center}
\end{table}

We now digress to discuss how the present work relates
to the standard ``improvement program" (see \eg the review~\cite{s}). 
In the study of the hadron spectrum,
only a single parameter (the bare quark mass) in the fermion action 
needs to be tuned. Once the correct continuum limit has been established,
attention is focused on eliminating those lattice artifacts that
vanish most slowly, that is, linearly with the lattice spacing.
However, in the calculation of weak matrix elements one has
to first establish the correct continuum limit.
This is very problematic with the standard fermion
methods because, due to the loss of full chiral symmetry, 
many subtraction coefficients must be tuned.  
Thus, having good chiral properties is of higher priority than 
the removal of any other lattice error. 
Also, in the massless-quark limit,
$O(a)$ lattice artifacts are automatically excluded
if chiral symmetry is maintained~\cite{msbk}.
In that sense, approaching the chiral limit using DWF
encompasses the standard improvement program as well.

Coming back to the new DWF action, 
since the new Wilson term $W_2$ starts off at order $p^4$,
the first lattice deviation from a relativistic (tree-level)
dispersion relation comes only from the kinetic term.
This is shown in the second column of Table~1. We observe that while
increasing $c_3$ from zero to $1/3$ improves the dispersion
relation, the opposite is true for $c_3>1/3$.
Although the error is formally of order $a^2$,
it might become significant if $c_3$ is too large. 
To gain some idea on the magnitude of the error 
consider, say, $p^2 \sim (400\; \mbox{\rm MeV})^2$, 
which is relevant for kaon physics, on a lattice with 
$a^{-1} \sim 2\; \mbox{\rm GeV}$. 
This means $a^2 p^2 \sim 1/25$. 
For the last two rows of Table~1, the effect is 2\% and 4\% respectively.

If next-next-nearest neighbors are also allowed
(still only in the same direction) one can obtain 
a smaller ${\rm max}\{\exp(-\a)\}$
while maintaining a vanishing $p^3$ term. Let
\beq
 D_{35} = \sum_\m \g_\m f_5(p_\m) + r\, W_3 - M \,,
\label{d35}
\eeq
where $W_n$ is defined in \seeq{wn} and
\beq
  f_5(p_\m) =  \sin(p_\m)
  \left[1 + {1\over 3}(1-\cos(p_\m))+c_5(1-\cos(p_\m))^2\right] \,.
\label{f5}
\eeq
Some values of ${\rm max}\{\exp(-\a)\}$ are given in Table~2.
Considering the last row ($c_5=50$), the deviation from Lorentz covariance
is at the level of $(50/4)(a^2 p^2)^2 \sim 2\%$ for the same $p^2$.
For $c_5=5$, the deviation is below 2\% up to $(700\; \mbox{\rm MeV})^2$.

\begin{table}[htb]
\begin{center}
\caption{${\rm max}\{\exp(-\a)\}$ for the operator $D_{35}$
and various values of $c_5$ and $r$.
The second column gives the first two terms
in the expansion of $f_5(p)$.
}
\begin{tabular}{ccccc}
\hline
$c_5$ &  $f_5(p)$ &   $r$ &   
$ 2 {\rm min}\{\cosh(\a)\}$    &    ${\rm max}\{\exp(-\a)\}$ \\
\hline
5  &   $p + 1{13\over 60}\, p^5$ &  2.21 &  17.74 &  0.0566  \\
10 &   $p + 2{14\over 30}\, p^5$ &  3.18 &  25.44 &  0.0394  \\
30 &   $p + 7{7 \over 15}\, p^5$ &  5.70 &  45.51 &  0.0220  \\
50 &   $p +12{7 \over 15}\, p^5$ &  7.48 &  59.84 &  0.0167  \\
\hline
\end{tabular}
\end{center}
\end{table}
The values of $\qpt={\rm max}\{\exp(-\a)\}$ in Table~2 are remarkably small.
Truly, the precise relation between $\qpt$ and
the non-perturbative $\qeff$ is not known. 
Since for ordinary DWF $\qpt=0.5$ while $\qeff \sim 0.8$,
one may hope that $\qpt$ provides at least an order-of-magnitude
estimate of $\qeff$. Even under this mild assumption,
it seems that the use of $D_{23}$ or $D_{35}$ 
may allow one to reduce $N_s$ a lot
while maintaining (or even improving) the quality of chiral symmetry
compared to the ordinary DWF action.

Replacing the four-dimensional part of the DWF action
by $D_{23}$ ($D_{35}$) approximately doubles (triples) 
the number of entries in the DWF matrix. 
Therefore one should expect an increase in the cost of simulations
by a factor of two (three) at fixed $N_s$. 
Our above considerations suggest that it may be possible to
compensate for this by working at smaller $N_s$
(at least in those cases where ordinary DWF require very large $N_s$).

In the continuum limit, all the DWF actions considered here support
a single quark (one chiral zero mode on each boundary) for $|1-M|<1$.
If the (generalized) Wilson term $W_n$ is employed, 
there is a four-quark zone
(corresponding to the doubler states in $P_\p$) at $|1-M+r\,2^n|<1$. 
An additional benefit of the new actions
is that the four-quark and the single-quark zones
are separated by a large gap (as a function of $M$)
which we expect to persist also in realistic simulations.

\section{CONCLUSION}

While the cost of simulating DWF is much bigger than that of
ordinary Wilson fermions, existing results show that this is
more than compensated by the much better quality of chiral symmetry.
Similarly, employing one or both of the modifications
proposed here might turn out to have enough benefits to outweigh
the extra cost of simulating the corresponding DWF actions.
In particular, it is hoped that these new actions may make it feasible 
to simulate DWF on coarse lattices ($a^{-1} \sim 1 \GeV$)
without loosing the good chiral properties.

\vspace{3ex}
\noindent {\bf Acknowledgements}
\vspace{2ex}

\noindent I thank T.\ Blum, N.\ Christ, M.\ Creutz, R.\ Edwards, 
M.\ Golterman, B.\ Mawhinney, P.\ Vranas and M.\ Wingate 
for useful comments.
Part of this research was carried out during a visit to
Brookhaven National Laboratory.
This research is supported in part by the Israel
Science Foundation.


\end{document}